\documentclass[apjl]{emulateapj}
\usepackage{hyperref}
\usepackage{color}
\hypersetup{colorlinks,
 citecolor=blue,
 linkcolor=blue}

\hypersetup{colorlinks,%
 citecolor=blue,%
 linkcolor=blue}
\definecolor{lightblue}{rgb}{.70,.95,1}

\def\aj{AJ}% Astronomical Journal
\def\apj{ApJ}% Astrophysical Journal
\def\apjl{ApJ}% Astrophysical Journal, Letters
\def\apjs{ApJS}% Astrophysical Journal, Supplement
\def\aap{A\&A}% Astronomy and Astrophysics
\def\aaps{A\&AS}% Astronomy and Astrophysics, Supplement
\def\mnras{MNRAS}% Monthly Notices of the RAS
\def\nat{Nature}% Nature
\def\pasp{PASP}% 
\def\araa{ARA\&A}
\def\physrep{PHYREP}
\def\cjaa{CIJAA}
\def\na{NewA}
\def\pasa{PASA}

\slugcomment{Draft Version \today}
\shorttitle{Dwarf-Dwarf merger}
\shortauthors{Paudel et al.}
\begin{document}
\title{A case study for a tidal interaction between dwarf galaxies in UGC 6741}

\author{Sanjaya Paudel\altaffilmark{1,2}, P. A. Duc  \altaffilmark{2},  C. H. Ree \altaffilmark{1}
  }
\affil{
$^1$ Korea Astronomy and Space Science Institute, Daejeon 305-348, Republic of Korea\\
$^2$ Laboratoire AIM Paris-Saclay, CEA/IRFU/SAp, 91191 Gif-sur-Yvette Cedex, France }

\altaffiltext{2}{Email: sanjpaudel@gmail.com}

\begin{abstract}
We present a case study of the tidal interaction between low mass, star-forming,  galaxies initially found exploring the Sloan Digital Sky Survey (SDSS) images and further analyzed with SDSS spectroscopy and  UV GALEX photometry.  With  a  luminosity of M$_{r}$ = $-$17.7 mag and exhibiting a prominent tidal   filament, UGC~6741  appears as a scale down version of massive gas--rich interacting systems and mergers.The stellar disk of the smaller companion, UGC~6741\_B, which is three times less massive, has likely been already destroyed.  Both  galaxies, which are connected by a 15~kpc long stellar bridge, have a similar oxygen abundance of 12+log(O/H)$\sim$8.3. Several knots of  star-forming regions are  identified along the  bridge, some with masses exceeding  $\sim$10$^{7}$  M$\sun$. The most compact of them, which are unresolved,   may evolve into globular clusters or Ultra Compact Dwarf galaxies (UCDs). This would be the first time progenitors of such objects are detected in  mergers involving dwarf galaxies. UGC~6741  has  currently the color and star formation properties of Blue Compact Dwarf galaxies (BCDs). However the analysis of its  surface photometry   suggests that the galaxy lies within the scaling relations defined by early-type dwarf galaxies (dEs).  Thus UGC~6741 appears as a promising system to study the possible  transformation of BCDs into  dEs, through possibly a merger episode. The frequency of such dwarf-dwarf mergers should now be explored.
\end{abstract}


\section{Introduction}
Since Halton Arp published an atlas of peculiar galaxies and \cite{Toomre72} reproduced their shape with numerical simulations of tidal encounters, mergers  have become key processes in the understanding of galaxy evolution.
Simulations show how  colliding disk galaxies are reshaped into  pressure-supported early-type bodies after the final merger. In a $\Lambda$CDM cosmology \citep{Spergel07}, which assumes that the assembly of large scale structure happens in a hierarchical fashion,   mergers  play a fundamental  role in the growth and evolution of galaxies \citep{Conselice09}. 

During the intermediate phases of interactions,  large scale tidal interactions trigger the formation of peculiar features like shells, streams, bridges and tails. The presence of such structures which is predicted by numerical simulations  is now frequently  observed in deep imaging surveys \citep{Duc11,Duc14b,Kim12,Struck99,vanDokkum05}. 

Numerous works have detailed the various phenomena occurring during galaxy interactions, including the formation in the collisional debris of substructures, like  Tidal Dwarf galaxies (TDGs)  \citep{Duc07}. However the vast majority of such studies focused on massive galaxies. Not much is known on tidal forces  generated by the encounter between low mass galaxies. 

It is common belief that,  having a shallow potential well,  low mass galaxies have  an evolution which is more driven by the large-scale environment than by  merging events. Dwarf galaxies exhibit a strong morphological segregation: the most evolved / oldest dwarf galaxies, i.e dwarf Spheroidal (dSph) or dwarf early-type (dE), are found in the group and cluster environments \citep{Kormendy09,Lisker09}, while  dwarfs with on-going  star-formation activity, such as Blue Compact Dwarf galaxies (BCDs, \citealt{Gil03,Papaderos96}) are mainly found in less dense environments. 
However, how precisely the environment contributes to the transformation  of star-forming dwarfs to anemic dEs is still a puzzle as several processes may play a role  \citep{Boselli06}.  Besides,  the mechanism that triggers the burst of star-formation in dwarf galaxies, particularly in BCDs, also remains a mystery.  Mergers, fly-by encounters or gas turbulence have been proposed \citep{Bekki08,Noeske01,Pustilnik01}.

Very recently, observational evidence for mergers between dwarf galaxies has  been growing \citep[e.g.][]{Amorisco14,Crnojevic14,Toloba14,Delgado12,Johnson13,Nidever13,Graham12,Penny12,Rich12,Geha05}. The possibility that dEs might be formed through mergers like the massive ellipticals has been proposed. If this is the case, one would expect the progenitors of dEs to exhibit the characteristics features of  mergers such as tidal tails.

The system presented here, UGC~6741,  was initially disclosed in a systematic eye inspection of tidal debris in the Sloan Digital Sky Survey (SDSS) images. It turned out to be  classified as a low surface brightness galaxy pair in \cite{Schombert98}. 
We report a multi-wavelength study of the system based on  archival optical images and spectra from the SDSS  DR7 release and UV-images from the Galaxy Evolutionary Explorer (GALEX) all sky survey \citep{Abazajian09,Martin05}.

 \begin{figure}
\includegraphics[width=9cm]{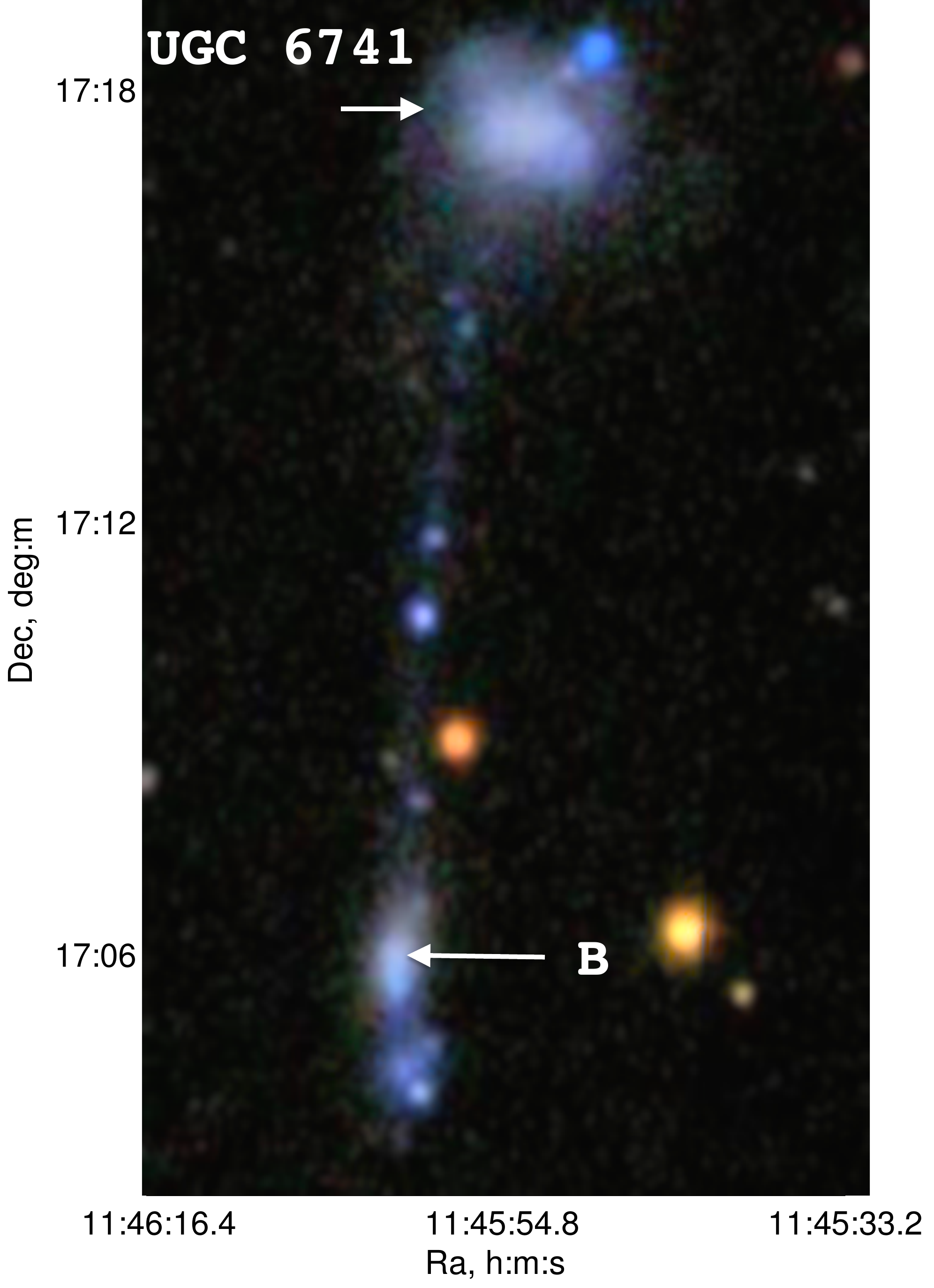}
\caption{True color $g+r+i$ SDSS image of  UGC~6741. Compact blue star-forming regions are distinctly visible in the tidal bridge between UGC~6741 and its companion disrupted galaxy, UGC~6741\_B. The 65" $\times$ 100" cutout image was queried from the SDSS DR7 sky server.}
\label{cfig}
\end{figure}

   \begin{figure}
\includegraphics[width=8.5cm]{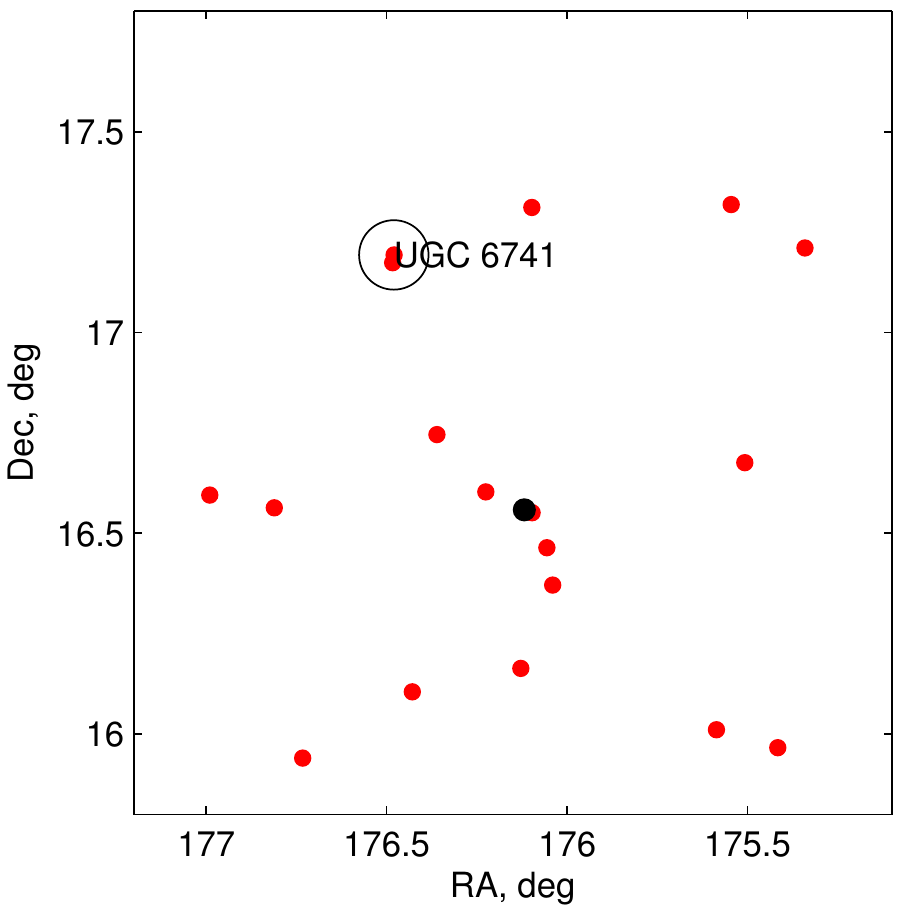}
\caption{Position of UGC~6741 (en-circled)  in the NGC~3853 group. The  central galaxy, NGC~3853, is shown with a  black dot and the other group members with  red dots. }
\label{grp}
\end{figure}

 \section{Data and Analysis}

\subsection{Target selection and location}

 In an our effort to search for tidal features around  low mass nearby galaxies, we visually inspected true color ($g+r+i$) SDSS images of nearly 40 000 galaxies located at   redshift below  0.02.
The  most prominent tidal feature  was  found in the galaxy pair, UGC~6741.
 With a luminosity M$_{r}$ = $-$17.7 mag,  UGC~6741 is slightly fainter than the two well known local group dwarf galaxies, the Large Magellanic Cloud (LMC) and NGC~4449 \footnote{Incidentally  these two latter galaxies are also involved in a tidal interaction.}. Its interacting companion, here after UGC~6741\_B,    has a luminosity of  M$_{r}$= $-$16.2 mag  \footnote{Throughout the paper, we assumed a distance to the galaxy of  D =  54.2 Mpc, which is the distance of the main galaxy group member, NGC~3853 provided by NED.}.

By chance,   UGC~6741 benefits from substantial  multi-wavelength data  in public archives, which  allowed  us to perform a thorough  analysis of its morphology,  chemical properties and stellar populations.

As shown in  Figure~\ref{cfig} and Figure~\ref{mfig}, a 15 kpc  long stellar bridge connects  the two galaxies along the North-South direction.    
 It hosts a number of  compact blue clumps -- the most prominent  ones are named objects A, C, D, E and F. Object F is located to the North of  UGC~6741, and either belongs to a secondary tail or to the bridge if the latter wraps around in the proximity  of the galaxy. Overall   the  system resembles classic mergers of massive gas-rich disk galaxies. 

 UGC~6741 is located on the outskirts of a group with an angular distance of 0.73 degree from the central galaxy, NGC~3853. In Figure~\ref{grp}, we show the group member galaxies located around a radius 700 kpc centered on NGC~3853   and with relative radial velocities  within $\pm$500 km/s. We used the NED database to carry out this search. The  difference in radial velocity between  NGC~3853 and UGC~6741 is less than 100 km s$^{-1}$.

\subsection{Imaging and Photometry}

\begin{figure*}
   \includegraphics[width=18cm]{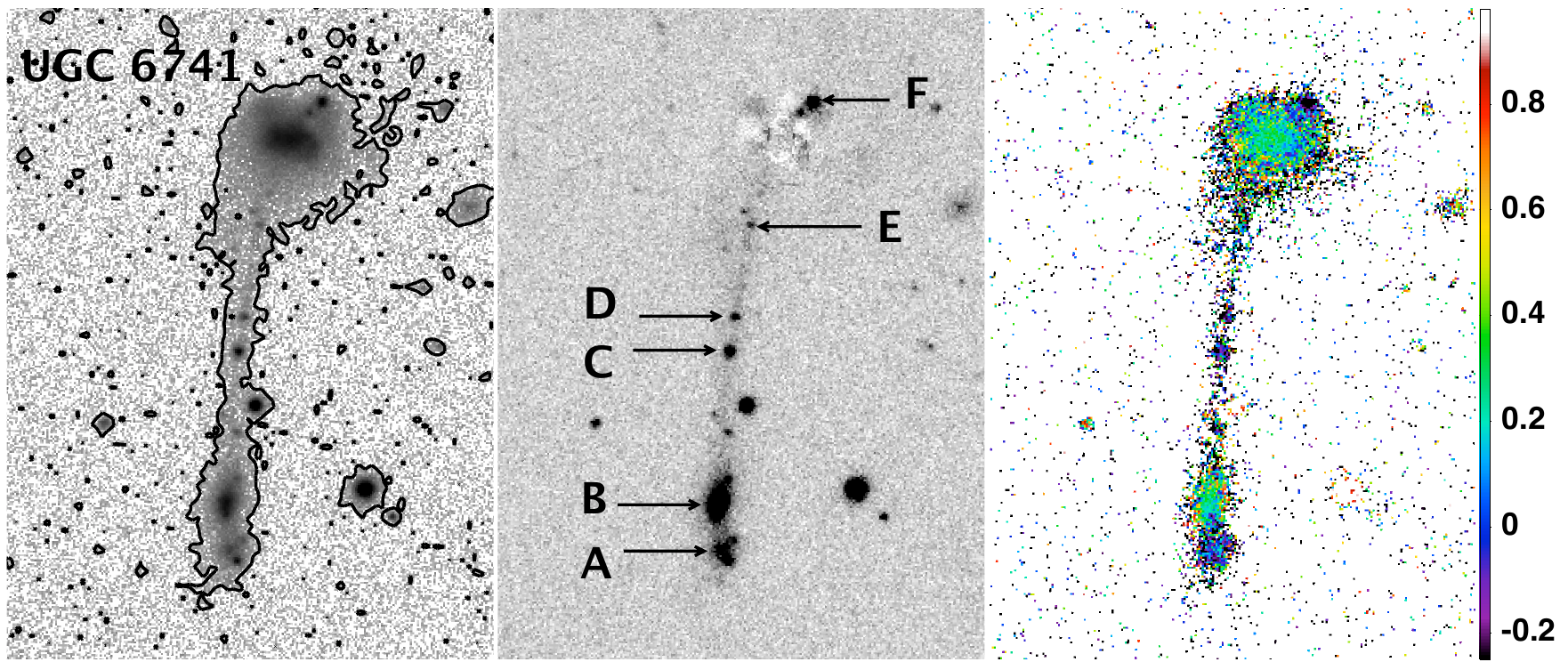}
  \caption{The interacting system, UGC~6741, as seen in the optical with the SDSS.  {\it Left:} co-added $g+r+i$ image with an arcsinh scaling. 
  The black contours represent the detection limit of 3$\sigma$  corresponding to a surface brightness level of $\sim$25.5 mag-arcsec$^{-2}$ in the r-band.  {\it Middle:} Residual image after having subtracted a model galaxy of UGC~6741.  Prominent star-forming clumps are named in alphabetical order starting from the bottom.  Note that  object B is likely  the main body of the  low-mass companion of UGC~6741 and is referred as UGC~6741\_B in the main text. 
 {\it Right:}   $g-r$ color map, with the scale in mag indicated to the right.  The field  of view of each image is 90" $\times$ 120". }
  \label{mfig}
   \end{figure*}

To perform a detailed image analysis, we retrieved archival images from the SDSS-DR7 database \citep{Abazajian09}. We used the $r$-band image as a reference, since it provides a higher signal to noise ratio on the tidal debris than the other bands. The seeing in this field is  0.9" (as measured from the $r$-band PSF).  In order to enhance the detectability of the bridge and the faint objects within it, a $g+r+i$ co-added image was made (see Fig.~\ref{mfig}, left), and a galaxy model subtraction has been carried out (see Fig.~\ref{mfig}, middle).  On such images, object B appears to be much more luminous,  extended and redder (see Fig.~\ref{mfig}, right) than the clumps A, C, D, E and F located along the tidal feature. As previously mentioned, B is most likely the main body of the disrupted  companion of UGC~6741.
The bluest clump, object E, is  located at the upper tip of the bridge and  embedded within the stellar halo of UGC~6741. Object A is somewhat irregular and  contains several compact sub-clumps.

\begin{figure}
\includegraphics[width=8.5cm]{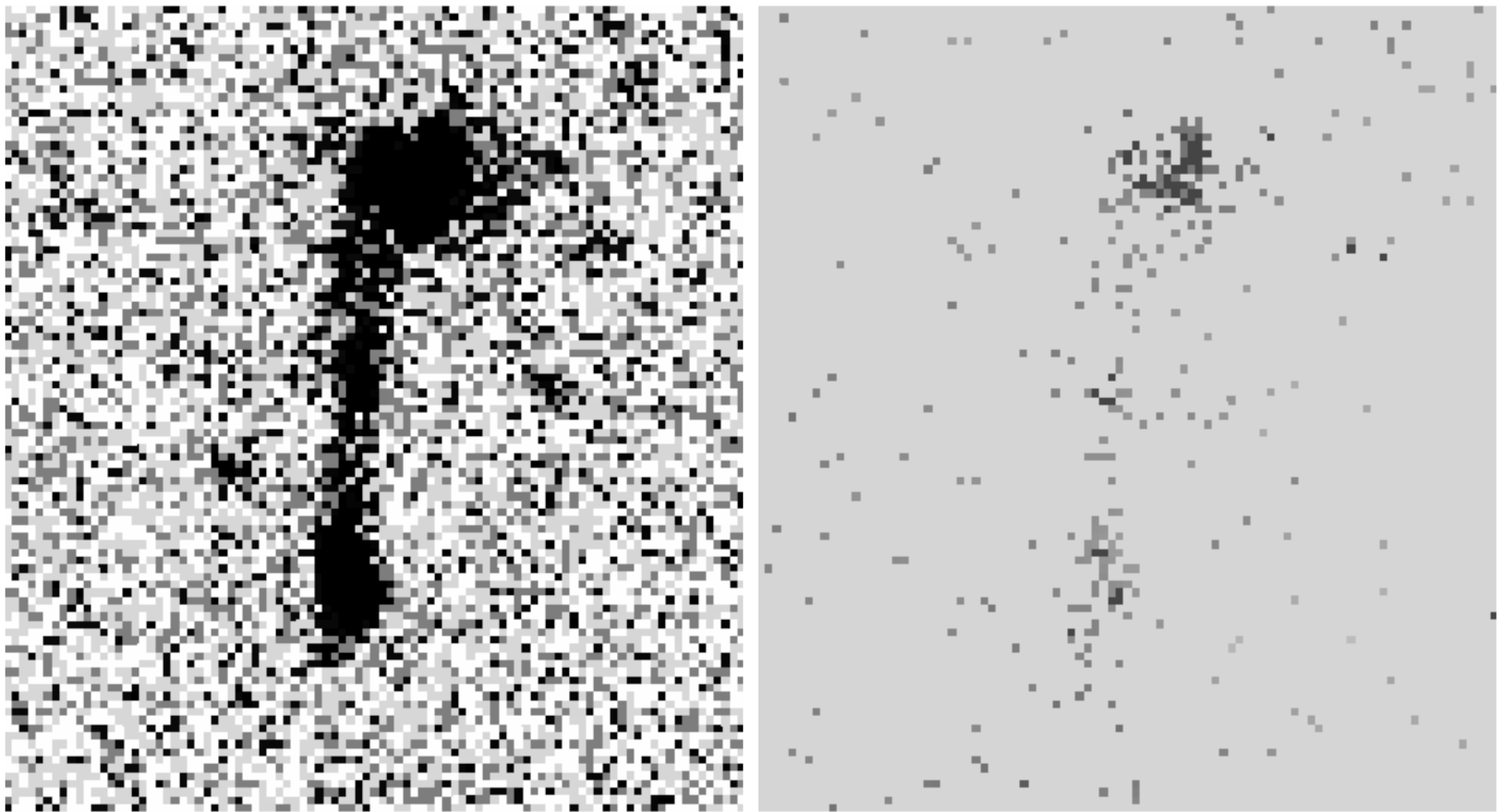}
\caption{GALEX all sky survey NUV (left) and FUV (right) images of UGC~6741.}
\label{galx}
\end{figure}

The  total luminosity of the clumps was determined from aperture photometry. The sky value has been estimated  in 5 independent regions  of 10$\times$10 pixel randomly chosen  around UGC~6741.  Their mean value was computed and subtracted from the measured fluxes.
We list the positions, absolute magnitudes and stellar masses of the clumps in Table \ref{ftb}. 
The derived magnitudes were  corrected for the Galactic extinction using \cite{Schlafly11}, but not for the internal extinction. 
We find that all of them  have $g-r$ color index less than 0.25 mag and the bluest, F, has a  $g-r$ color index as low as -0.03 mag. Their blue color and detection in the UV by GALEX (see Fig \ref{galx}) is totally consistent with the hypothesis that they are star-forming regions.
The stellar masses were derived from the SDSS$-r$ band magnitude with a mass to light ratio tabulated by \cite{Bell03} and appropriate to the observed  $g-r$ color.  These  values are most likely upper limits.

 \begin{table*}
 \caption{ Global properties of the system}
\begin{tabular}{lccccclccc}
\hline
Galaxy &Ra & Dec & Mr & v$_r$ & M$_{*}$ & SFR(m$_{FUV})$ & 12 + log(O/H)  & R$_{e}$ & $<\mu>_{r}$\\
\hline
 & $^{o}$ &  $^{o}$ & mag & km s$^{-1}$ & M$_{\sun}$ & M$_{\sun}$ yr$^{-1}$ & dex & kpc & mag arcsec$^{-2}$\\
\hline
UGC~6741       & 176.4793 & 17.1923  &  $-$17.74$\pm$0.01  &      3390  &  $6.0 \times 10^{8} $    &  0.09(17.96$\pm$0.08)   &  8.3 & 1.2 & 21.71 \\ 
UGC~6741\_B  & 176.4827 & 17.1732  &  $-$16.16$\pm$0.01  &      3300  &  $ 1.9 \times 10^{8} $     &  0.03 (19.01$\pm$0.12)$^{*}$  &  8.3 &   -- & -- \\ 
\hline
\end{tabular}
\label{ptb}
{\\
\\
 The SFR is estimated from the FUV magnitude  given in the parentheses.  A galactic extinction correction has been applied using the formula A$_{FUV}$ = 8 $\times$ E(B-V) with E(B-V) = 0.2.  The stellar mass is derived from the $r-$band luminosity and the mass to light ration obtained from \cite{Bell03} for the color $g-r$. The value of the Oxygen abundance, 12 + log(O/H),  is derived with the  O3N2 method which has a typical systematic error of  0.2 dex. \newline
$^{*}$ The FUV flux and derived SFR includes the star forming region A, see Fig \ref{mfig}.}
\end{table*}

The extended Clump A, located at the  tip of the tidal feature,  is the brightest star-forming region. The fairly compact regions  C, D and and E  are actually likely wrongly  classified as stars in the SDSS catalog. Our measured FWHM for these  knots are well consistent with the median FWHM of foreground stars, i.e. 0.9" and in the absence of available   radial velocities, we cannot  totally exclude chance superposition of  foreground blue stars or background compact objects. However, the remarkable alignment of these objects along the stream and their similarity in color (see Figure~\ref{cfig} \& \ref{mfig}) are strong hints for a real physical association with the system.

The interacting dwarf galaxies and associated tidal debris are  clearly detected in the NUV, and barely detected in the FUV by the GALEX all sky survey (see Fig.~\ref{galx}).  Since the GALEX images have a spatial resolution of only 5", the star-forming regions are not resolved individually in the UV images. The  star formation rates, derived from the FUV fluxes applying a  Galactic extinction correction from  \cite{Schlegel98}, and using the calibration of \cite{Kennicutt98},   are given in Table \ref{ptb}.

\begin{figure}
\includegraphics[width=8.5cm]{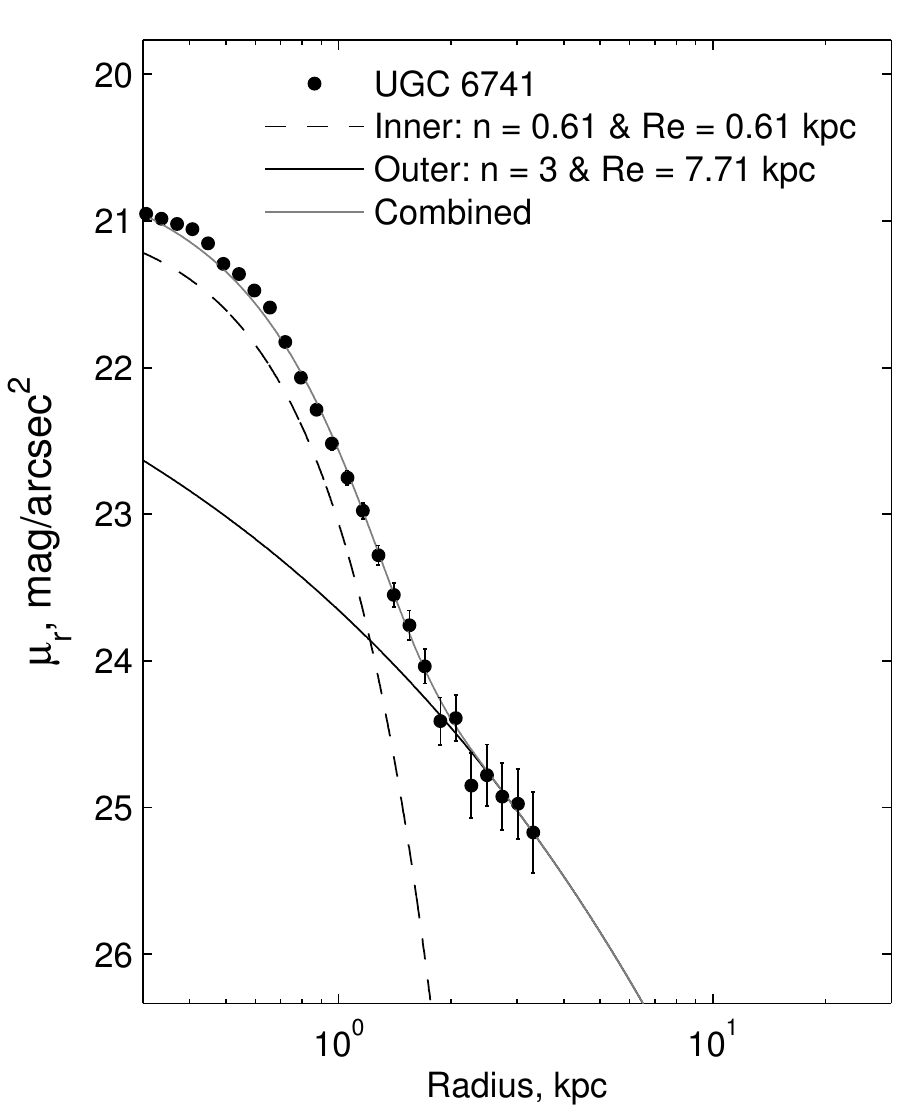}
\caption{Major axis radial profile of UGC~6741, where the observed profile is fitted with two component S\'ersic model.  }
\label{prof}
\end{figure}

With the help of the IRAF $ellipse$ task, we performed a surface photometry analysis of  UGC~6741. First, we extracted the galaxy major-axis light profile from the $r-$band image. In doing so, the centre and position angle of the ellipse were held fixed and the ellipticity was allowed to vary. The centre of the galaxy was calculated using the task $imcntr$ and input ellipse parameters were determined using the several iterative runs of the $ellipse$ task. The derived $r-$band major axis light profile of UGC~6741 is shown in Figure~\ref{prof}. With the help of $\chi^{2}$-minimization scheme, we obtained a best fit model where the observed profile is decomposed into the two component S\'ersic functions. The inner and outer components are best fitted with the S\'ersic functions of n = 0.61 and 3, respectively.

Since the observed light profile of UGC~6741 is better represented by a multi component S\'ersic function the derivation of the effective radius and effective surface brightness is not obvious.  Therefore, we estimated these parameters with a non-parametric approach. Using the similar procedure followed by \cite{Janz08}, we first calculated the Petrosian radius and the total flux was measured within  two Petrosian radius. Note, that contrary to   \cite{Janz08},  we do not correct for the the missing flux estimated  by \cite{Graham05}. Indeed this correction is very small for dwarf galaxies \citep{Chen10}. The derived values of the structural parameters of UGC~6741 are given in Table \ref{ptb}. We find that the mean surface brightness within the half-light radius is 21.7 mag arcsec$^{-2}$ in the SDSS $r-$band. This is in fact a rather low value of the surface brightness of a low-surface brightness galaxy, as described in \cite{Schombert98}.

   \begin{table}
   \caption{Photometry of the  star forming tidal clumps}
   \setlength{\tabcolsep}{0.06cm}
\begin{tabular}{lcccccc}
\hline
 & Ra & Dec  & g  & r & i & M$_{*}$  \\
  &$^{o}$ & $^{o}$ & mag & mag  & mag& 10$^{8}$M$\sun$  \\
\hline
A & 176.4821 & 17.1704 & 18.46$\pm$0.02 & 18.52$\pm$0.03  & 18.39$\pm$0.04 & 0.35  \\ 
B & 176.4827 & 17.1732 & 17.76$\pm$0.02 & 17.51$\pm$0.02  & 17.37$\pm$0.02 & 1.94  \\ 
C & 176.4820 & 17.1810 & 19.42$\pm$0.03 & 19.55$\pm$0.04  & 19.33$\pm$0.05 & 0.11  \\ 
D & 176.4817 & 17.1827 & 20.51$\pm$0.05 & 20.73$\pm$0.08  & 20.44$\pm$0.10 & 0.03  \\ 
E & 176.4808 & 17.1875 & 20.83$\pm$0.06 & 21.07$\pm$0.10  & 20.59$\pm$0.10 & 0.02  \\ 
F & 176.4775 & 17.1940 & 18.05$\pm$0.01 & 18.08$\pm$0.02  & 18.23$\pm$0.03 & 0.57  \\ 
U & 176.4793 & 17.1923 &16.05$\pm$0.02 & 15.93$\pm$0.02  & 15.77$\pm$0.03  & 6.02  \\   
\hline
\end{tabular}
{\\
\\
The galactic extinction is corrected with \cite{Schlafly11}. Stellar masses are converted from the $r-$band magnitude using the mass to light ratio from \cite{Bell03}, i.e Log(M/L) = $-$0.306+1.097($g-r$)}
\label{ftb}
\end{table}

Besides, we performed the aperture photometry on the entire system. For this, we first manually masked all non-related objects  and selected a large aperture that includes the entire system. The total $r-$band luminosity is m$_{r}$ = 15.53 $\pm$0.01 and $g-r$ color index is 0.1 $\pm$0.01 mag. The FUV and NUV luminosities are 17.21 $\pm$07 and 17.06 $\pm$03 mag, respectively. Using the same conversion factors as for the individual sub-structures, we derived a total stellar mass  M$_{*,\,total}$ = 8.3$\times$10$^{8}$ M$\sun$ and a total star-formation rate SFR$_{total}$ =  0.18 M$\sun$ yr$^{-1}$.

  \subsection{Gas content}
Atomic  hydrogen (HI) 21-cm radio data are available in the Hyperlyda archive \citep{Paturel03}.   Observations of  UGC~6741 were made with the Arecibo single dish telescope. The large beam size, $\sim$3' \citep{Lu93} , covered  the entire interacting system. The cataloged archival HI parameters are listed in Table \ref{hi}.  We computed a total HI mass of  8.7$\times10^{8}$  M$_{\sun}$, and an inferred  HI mass to blue luminosity ratio, M$_{HI}$/L$_{B}$,  of 0.65 M$_{\sun}$/L$_{\sun}$ \footnote{The B-band luminosity was derived from the $g-$band luminosity using the empirical formula of \cite{Lupton05}, i.e  B = $g$+0.313($g-r$) + 0.227 }.
  This number is similar to the average value of M$_{HI}$/L$_{B}$ for the sample of BCDs studied in \cite{Huchtmeier05} but significantly lower than in  typical isolated low-surface brightness dwarf galaxies  \citep{Pustilnik11}. We estimated the expected HI mass of UGC~6741 using an empirical relation between the HI mass and the diameter \citep{Gavazzi05},  $M_{HI ref} = a+b log(d)$. For this, we used the isophotal diameter provided by  NED and adopted the constants a = 7.00 and b = 1.88 for Scd-Im-BCD type galaxies. The expected HI mass is less than half of the observed HI mass in the whole system.

The HI spectrum shows a single velocity peak  \citep[see Fig. 1 in ][]{Lu93} that may suggest the absence of a kinematically distinct HI component for each galaxy. Note, however, that the observed velocity width is similar to the difference in radial velocity measured from the optical spectra of each galaxy, i.e. 76 km$^{-1}$. 

\begin{table}
\caption{HI gas content}
\begin{tabular}{l|l}
\hline
\hline
HI heliocentric velocity & 3324 km/s\\
HI Flux  & 1.26 Jy-km/s $\pm$10\%\\
HI mass & 8.7 $\times$10$^{8}$\\
Mean velocity width & 79.89 $\pm$1.2 km/s\\
M$_{HI}$/L$_{B}$  & 0.65 \\
\hline
\end{tabular}
\label{hi}
\end{table}

\subsection{Optical spectroscopy}
\begin{figure*}
 \includegraphics[width=18cm]{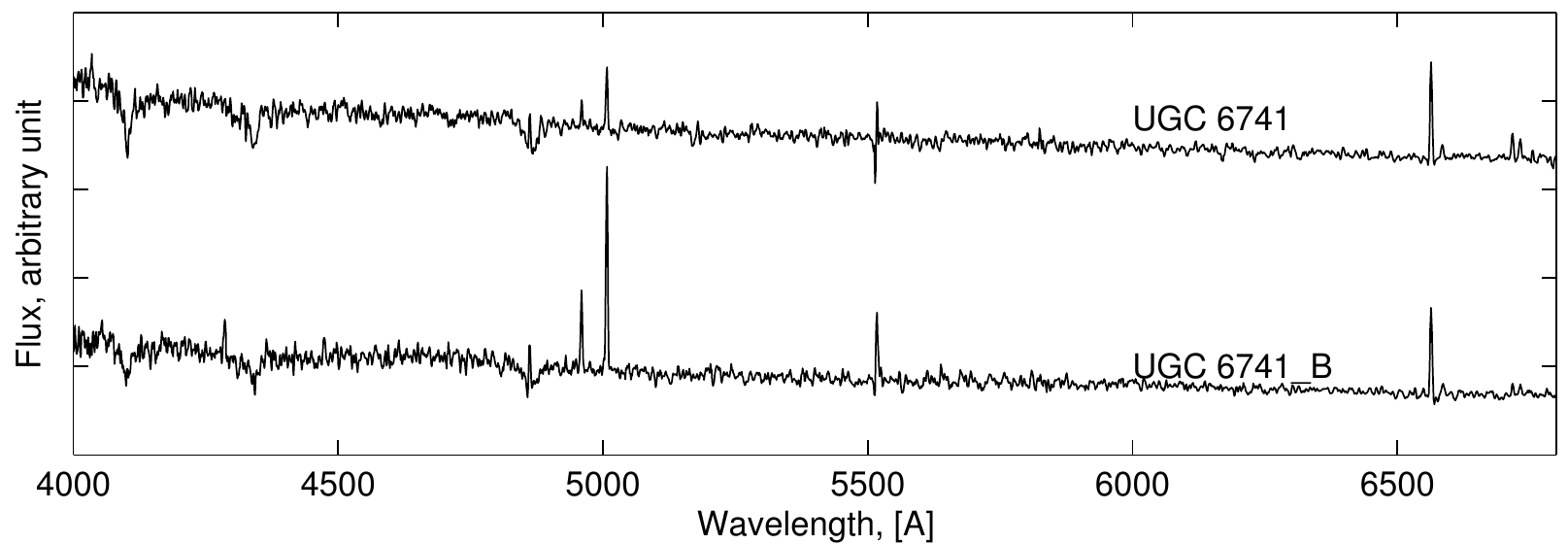}
 \caption{SDSS optical spectra of UGC~6741 and its companion UGC~6741\_B. The observed spectra are shifted to rest frame wavelength and smoothed with a 3 pixel gaussian kernel.}
 \label{spec}
\end{figure*}

The optical spectra of both UGC~6741 and UGC~6741\_B  were queried from the SDSS archives (see Fig.~\ref{spec}). They exhibit the  emission lines typical of HII regions as well as   relatively strong absorption  for the early  Balmer lines H$\delta\, ,\gamma\,$ and $\,\beta$.
The emission line fluxes were measured after subtracting the stellar absorption features  using the publicly available code  GANDLF \citep{Sarzi06} and the stellar templates of  \cite{Tremonti04}. 
 
The extinction coefficient, E(B-V), derived from the Balmer decrement H$\alpha$/H$\beta$  is nearly equal to zero for UGC~6741 and even below the standard value of 2.86 for UGC~6741\_{B}. This  may indicate a poor subtraction of the absorption at H$\beta$ due to the low signal to noise ratio. 

The H$\alpha$ equivalent widths -- 7.1 \AA{} and 8.7  \AA{} for UGC~6741 and UGC~6741\_B respectively --  are relatively small compared to the typical BCD's H$\alpha$ equivalent widths \citep{Gil03}.
 
Oxygen abundances, 12+log(O/H), were estimated with the  two  methods described, among others,  by \cite{Marino13}, i.e the so--called N2 and O3N2 methods. The N2 method only considers the line ratio between H$\alpha$ and [NII]  while the O3N2 method uses a combination of the line ratios H$\alpha$/[NII] and [OIII]/H$\beta$. We obtained 12+log(O/H) = 8.3(8.3) and 8.3(8.2) dex from the N2(O3N2) method for  UGC~6741 and  UGC~6741\_B, respectively. These values are in the range of typical  BCDs \citep{Gil03,Vaduvescu07}. 
The systematic error of the methods is 0.2 dex.

\section{Discussion and Conclusion}
We have presented a case of a dwarf-dwarf merger in a group environment. Remarkably our multi-wavelength study, based on imaging and spectroscopy,  could be carried out with data solely acquired from publicly available archives.

We discuss below the  origin of its  prominent star--forming tidal bridge and speculate on the future evolution of  UGC~6741.

 \subsection{Nature of the tidal features and its star-forming regions}
\begin{figure}
 \includegraphics[width=8.5cm]{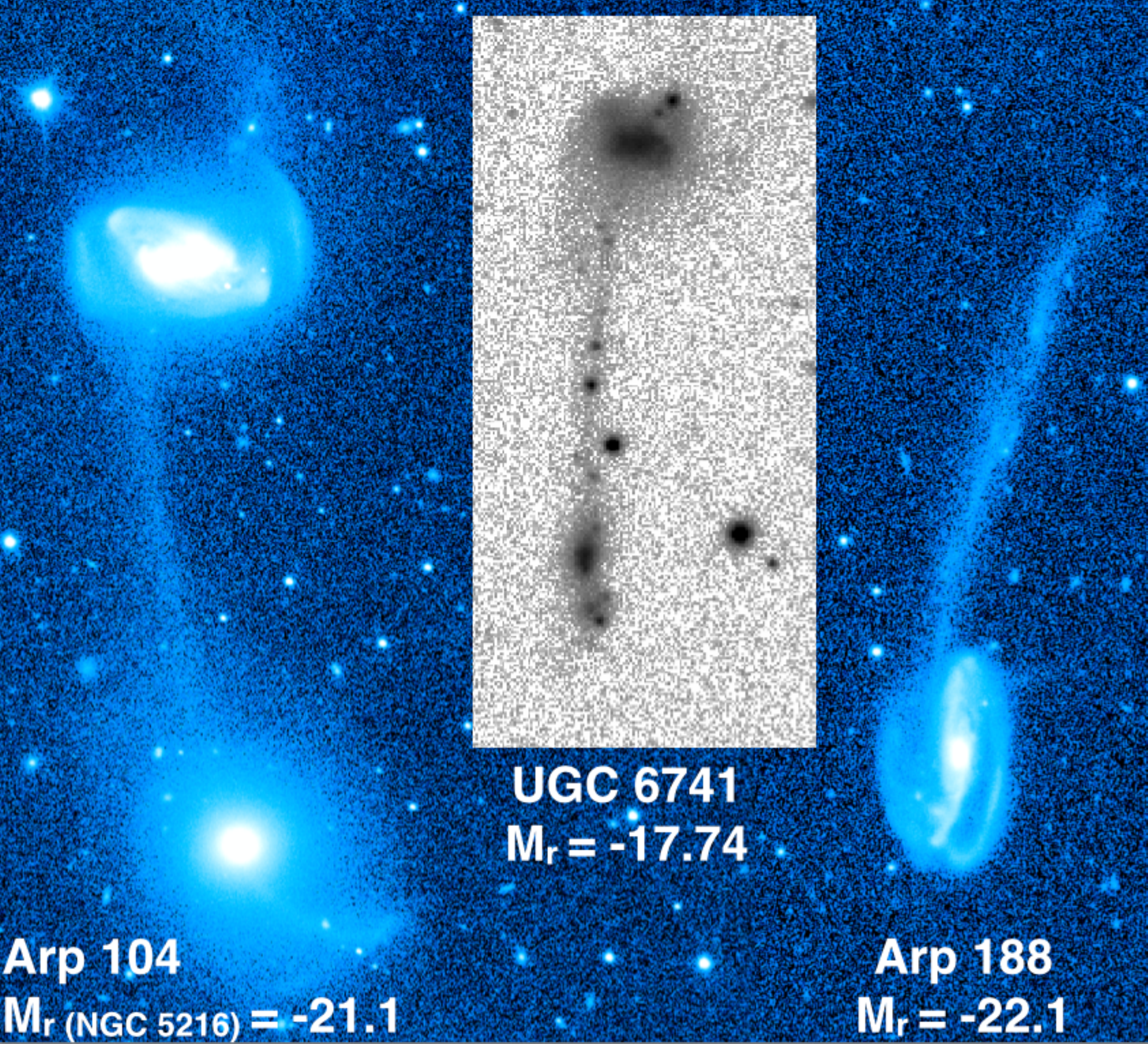}
 \caption{Direct comparison between Arp~104 and Arp~188 (in blue scale) with UGC~6741 (in grey-inverted scale). The optical $r-$band images for Arp~104 and Arp~188 were also obtained from the SDSS. }
 \label{comp}
\end{figure}

 UGC~6741 is noteworthy for the presence of a tidal bridge hosting  knots of star-formation, and in that respect resembles systems involving massive colliding  galaxies, such as   Arp~104  \citep{Gallagher10} or Arp~188 (UGC~10214), also known as the Tadpole galaxy.
  A comparison between  these systems is made in Figure~\ref{comp}. 
 
Like UGC~6741, Arp~104 exhibits a prominent tidal bridge, with hints that part of the tidal material wraps around one of the interacting galaxies (NGC~5218,  to the North). For this system however, there is no evidence for the presence of star--forming regions along the bridge. 
 
 Like in UGC~6741, the long single tail that emanates  from Arp~188 hosts multiple compact knots of star formation. Some of its  young massive star cluster  are as massive as $\sim$$10^{6}$ M$_{\sun}$ \citep{Tran03,Grijs03,Jarrett06} and were most likely formed in situ in the tail out of  gaseous material expelled from  UGC 10214 during a dynamical interaction  with a hidden or already destroyed companion. 
 
 Given its shape, the stellar filament  of the system studied here is undoubtedly of tidal origin.  Is  it however  as for  Arp~104 a bridge connecting two  interacting galaxies, namely UGC~6741 and  UGC~6741\_B, or a single tail like in Arp~188, with UGC~6741\_B being  a  Tidal Dwarf Galaxy
 instead of a pre-existing object?

Interestingly, the gas phase metallicity derived from the SDSS archival spectra reveals within the errors no difference in Oxygen abundance between UGC~6741 and UGC~6741\_B although the latter is three times less massive. The mass-metallicity relation would predict an abundance of about 0.2 dex lower. 

Similar metallicities between the parent galaxy and its tidal dwarf are instead expected as the latter is made from material pre-enriched in the former \citep{Duc07,Duc14}. This may in principle argue in favor of the TDG hypothesis for UGC~6741\_B. 
This is however without taking into account the rather large intrinsic scatter of the M-Z relation of 0.2 dex and of the systematic uncertainty in the abundance determination of the same order.

Besides, the existence of a prominent stellar continuum  (see Fig.~\ref{spec})  suggests that the object has a significant fraction of old stellar populations and  the presence of  strong Balmer absorption lines is a signature of star-formation over an extended period of time. The  TDGs  so far discovered are rather characterized by the overall dominance of young stars  and of an on-going instantaneous starburst. This  suggests that UGC~6741\_B is a pre-existing dwarf, which is interacting with UGC~6741. 
Therefore, the system would in fact be rather a scale down version of  Arp~104, with an overall  luminosity  $\sim$4 mag fainter.

Evidence of in situ star-formation occurring in  gas rich collisional debris has been reported in numerous massive interacting galaxies, including the Tadpole galaxy discussed above. Such regions are  believed to be also a nursery of   super star clusters \citep[e.g.][]{Grijs07,Mello08,Peterson09,Fedotov11}. 
Idealized numerical simulations of mergers reproduce the formation of massive and compact super star-clusters in tidal tails  \citep{Bournaud08,Renaud15}
and predict they might be the progenitors of globular clusters, provided they survive internal feedback and external tidal shear. 
The most massive and extended  of them may become independent tidal dwarf galaxies.

The same phenomena seems to also apply to dwarf-dwarf major mergers. The bridge of UGC~6741 hosts at at least four distinct knots of star-formation.  They are as massive as $\sim$10$^{7}$ M$\sun$;  with stellar mass density reaching   $\sim$10$^{8}$ M$_{\sun}$ kpc$^{-1}$
\footnote{Assuming  sizes equal to the observed FWHM}, they could evolve into  globular clusters  or even ultra compact dwarf galaxies. 
Their parent galaxies being low-mass systems with shallow potential well, one may speculate that they will survive longer than in the environment of systems involving massive merging galaxies.

 \subsection{Dwarf-Dwarf merger}
\cite{Lucia06} show that the statistical probability of merger of galaxies decreases towards the low mass regime. Our study of UGC~6741, however, proves  that dwarf-dwarf tidal interactions and mergers  occur in the nearby Universe. How frequent are they? 
On the one hand, exploring the  Millennium cosmological Simulation, \cite{Moreno13} found that binary  mergers in isolation are very rare, and claimed that satellite-satellite mergers should play a higher role than so far anticipated. 
On the other hand, \cite{Klimentowski10} used the Constrained Local Universe Simulation (CLUES) to conclude that interactions between satellites are unlikely. In a group environment, mergers between  sub-haloes are predicted to occur only before they entered the host halo.  
Interestingly,   UGC~6741 is itself located in the very  outskirts of a group.

In the real Universe,  reported cases of  possible dwarf-dwarf mergers have increased  recently \citep[e.g.][]{Amorisco14,Delgado12,Penny12,Rich12}. 
Distorted HI morphologies and the presence of gaseous tails around starbursting  dwarf galaxies  have also been attributed to mergers  \citep[e.g][]{Johnson13,Nidever13}. 
However, direct evidence for an on-going tidal interaction between objects of roughly similar masses remains  rare.  \cite{Rich12} found that  the nearby  Magellanic irregular galaxy NGC~4449 makes a pair with a tidally disrupted dwarf galaxy. This interaction will lead to a 1:50 merger, whereas  UGC~6741 corresponds to a 1:3 merger.
In the world of massive galaxies, this would be considered as a major merger.

\subsection{Evolution of dwarf galaxies}
Several physical properties of UGC~6741, i.e. color, metal content and star-formation rate, are fairly similar to typical blue compact dwarf galaxies and there is little doubt that its star formation activity  is affected, if not triggered,  by the interaction. Once the burst is terminated and the galaxies have merged, which type of dwarfs will UGC~6741 resemble? Its position on the scaling relations between  structural parameters can give clues on its future evolution.

\begin{figure}
\includegraphics[width=8.5cm]{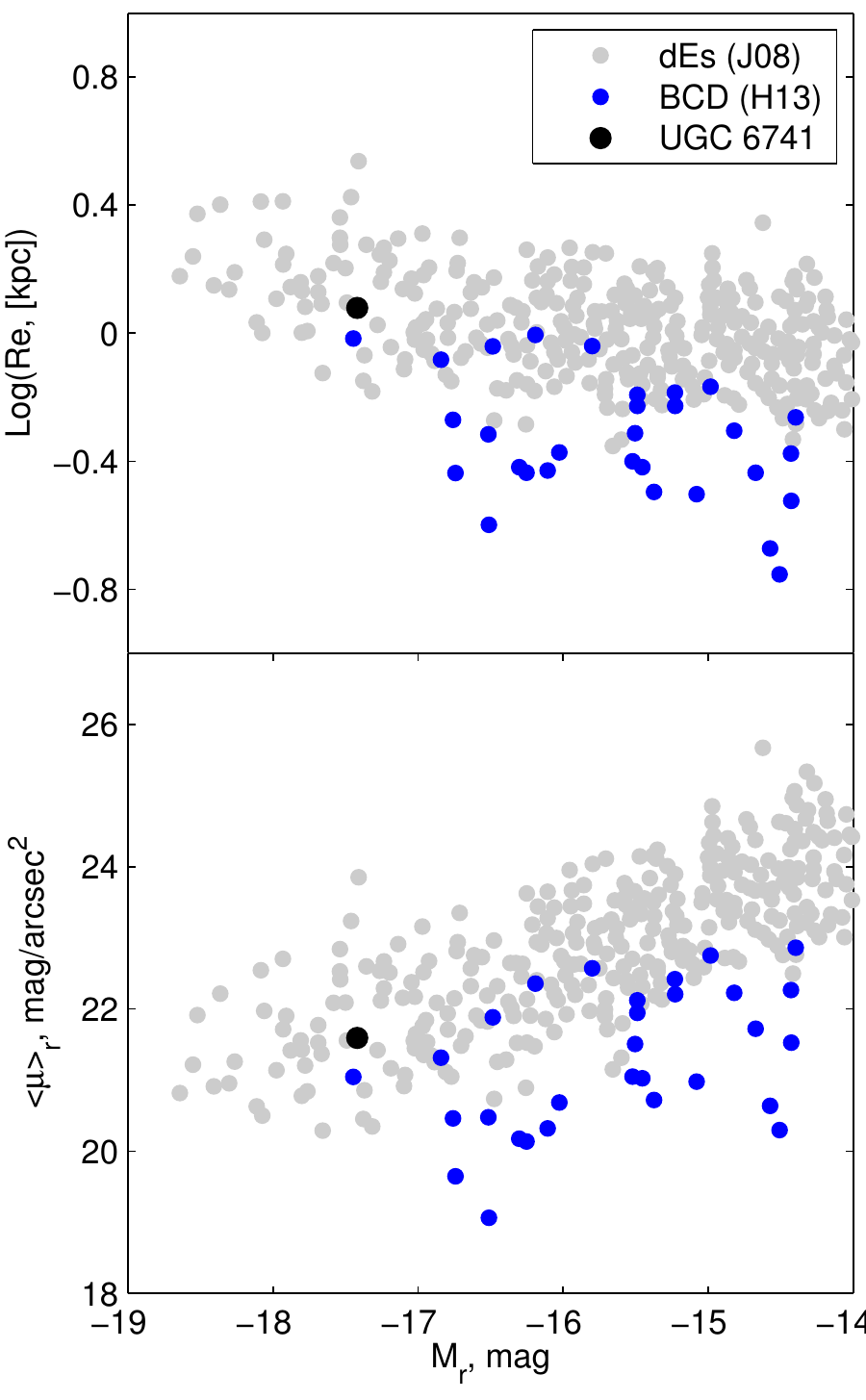}
\caption{Scaling relations for dwarf galaxies. The two well established relations, effective radius vs luminosity, Log(R$_{e}$) vs M$_{r}$ (top panel)  and surface brightness vs luminosity, $<\mu>_{r}$ vs M$_{r}$ (bottom panel)  are shown for a  sample dEs  (gray dots, from \cite{Janz08})  and BCDs (blue dots, from \cite{Meyer13}).  UGC~6741 is shown with the black solid circle.}
\label{scf}
\end{figure}

In Figure~\ref{scf}, we  compare its properties  with that of  samples of non-interacting  dwarfs. As a  comparison sample, we used the Virgo Cluster BCDs and dEs  from \cite{Meyer13} and \cite{Janz08} respectively.

 BCDs occupy a position distinct from dEs in both M$_{r}$ - $<$$\mu$$>$ and M$_{r}$ - Re scaling relations. In particular, they  have a higher surface brightness than dEs of  similar magnitude. UGC~6741  which has a total luminosity similar to the brightest BCDs in the comparison sample lies  within the locus of dEs. 
 \cite{Meyer13} proposed  a formation channel of some dEs through a BCD phase, an evolution scheme that UGC~6741  may also follow.
Yet,   given its current star formation rate of 0.18 M$_{\sun}$ yr$^{-1}$ and the mass of its gas reservoir estimated from the HI data,  its gas depletion time is high   ($>$3 Gyr).
Thus additional processes other than star-formation are needed to remove the gas reservoir and make UGC~6741  a gas-poor red and dead galaxy.  \\

Therefore, UGC~6741 appears as a scale down version of gas--rich massive interacting systems: like the latter, it hosts a prominent long tidal filament hosting young super star clusters that will possibly evolve into globular clusters or ultra compact dwarf galaxies. A further comparison study would require high resolution  optical imaging and spectroscopy  and to acquire HI/CO gas maps, in additional to the wealth of multi-wavelength data already available in the archives. 
Such observations would give clues on the future evolution of the system. Whether gas--rich dwarf-dwarf mergers  form dwarf elliptical galaxies the same way as   spiral-spiral mergers form massive ellipticals is still an open question. 
 Cosmological simulations have so far given somehow contradicting results on the merger probability in the dwarf population. 
A statistical census of tidally interacting dwarfs should now be carried out. This was a motivation for us to systematically inspect large scale imaging surveys such as the SDSS or the NGVS in the Virgo Cluster. Results of this statistical analysis will be presented in future papers (Paudel et al in preparation).

\acknowledgements
We thank the referee, Prof. Curtis Struck, for helpful comments which improved the paper.\\
This study is based on the archival images and spectra from the Sloan Digital Sky Survey (the full acknowledgment can be found at http://www.sdss.org/collaboration/credits.html) and  NASA Galaxy Evolution Explorer (GALEX). Funding for the SDSS has been provided by the Alfred P. Sloan Foundation, the Participating Institutions, the National Science Foundation, the U.S. Department of Energy, the National Aeronautics and Space Administration, the Japanese Monbukagakusho, the Max Planck Society, and the Higher Education Funding Council for England. The SDSS Web Site is http://www.sdss.org/. GALEX is operated for NASA by the California Institute of Technology under NASA contract NAS5-98034. We also acknowledge the use of NASA's Astrophysics Data System Bibliographic Services and the NASA/IPAC Extragalactic Database (NED).
  
\newpage

\newcommand{\adsurl}[1]{\href{#1}{ADS}} 
\providecommand{\url}[1]{\href{#1}{#1}} 
\providecommand{\newblock}{}

\end{document}